\documentclass[aps,twocolumn,showpacs,amsmath,amssymb,pra,superscriptaddress,floatfix,longbibliography]{revtex4-1}
\usepackage{braket}
\usepackage{amsmath}
\usepackage{amssymb}
\usepackage{mathtools}
\usepackage{graphicx}
\usepackage{dcolumn}
\usepackage{bm}
\usepackage{multirow}
\usepackage{leftidx}
\usepackage{color}
\usepackage{float}

\usepackage{amsfonts}
\usepackage{eufrak}
\usepackage[german,english]{babel}

\begin{document}
\title{
Theory of long range interactions for Rydberg states attached to
hyperfine split cores}
\author{F. Robicheaux}
\email{robichf@purdue.edu}
\affiliation{Department of Physics and Astronomy, Purdue University, West Lafayette,
Indiana 47907, USA}
\affiliation{Purdue Quantum Center, Purdue University, West Lafayette,
Indiana 47907, USA}
\affiliation{Department of Physics, University of Wisconsin-Madison, 1150 University Avenue, Madison, Wisconsin 53706, USA}
\author{D.W. Booth}
\affiliation{Department of Physics, University of Wisconsin-Madison, 1150 University Avenue, Madison, Wisconsin 53706, USA}
\author{M. Saffman}
\affiliation{Department of Physics, University of Wisconsin-Madison, 1150 University Avenue, Madison, Wisconsin 53706, USA}

\date{\today}

\begin{abstract}
The theory is developed
for one and two atom interactions when the atom has a Rydberg electron
attached to a hyperfine split core state. This situation is relevant for
some of the rare earth and alkaline earth atoms that have been proposed for
experiments on Rydberg-Rydberg interactions. For the rare earth atoms,
the core electrons can have
a very substantial total angular momentum, $J$, and a non-zero nuclear spin,
$I$. In the alkaline earth atoms there is a single, $s$, core electron whose spin
can couple to a non-zero nuclear spin for odd isotopes.
The resulting hyperfine splitting of the core state can lead to substantial
mixing between the Rydberg series attached to  different thresholds. Compared
to the unperturbed Rydberg series of the alkali atoms, the
series perturbations and near degeneracies from the different parity states
could lead to qualitatively different behavior for single atom Rydberg properties
(polarizability, Zeeman mixing and splitting, etc) as well as Rydberg-Rydberg
interactions ($C_5$ and $C_6$ matrices). 

\end{abstract}


\maketitle

\section{Introduction}

The structure and interactions of atoms excited to Rydberg states have 
been intensively studied for many years. Detailed experimental measurements
of Rydberg properties were initially  performed with alkali and alkaline earth
atoms\cite{TFG}. More recently there has been a growing
interest in the use of rare earth atoms, primarily lanthanides, for
experiments with degenerate quantum
gases\cite{LBY,LBL,AFM,FMA,SWB},
and for quantum information\cite{SM1}. Alkaline earth atoms are also
the subject of increased interest for quantum information
applications\cite{DBY,SKY}. The availability of Rydberg
state mediated potentials provides a tunable experimental control
parameter for studies of long range interactions and entanglement.
Several works have proposed incorporating Rydberg interactions in
experiments with alkaline
earth\cite{MMN,TD1,TD2,GMB,KLH} and
lanthanide\cite{SM1} atoms.

The Rydberg structure of these multi-electron atoms is substantially
more complex than for single electron alkali atoms. The standard
theoretical technique used to quantitatively describe these atoms
is multichannel quantum defect theory (MQDT)
as presented, for example, in Ref.~\cite{AGL}. Several recent works have used MQDT
to calculate the interaction potentials between Rydberg excited alkaline earth
atoms\cite{VJP,VJP2}. The combination of multiple Rydberg
series and a hyperfine split core state can lead to mixing between Rydberg
series attached to different thresholds leading to additional complexity. 
Hyperfine structure is present in alkaline earth and lanthanide isotopes with an 
odd number of nucleons and thereby a nonzero nuclear spin. These isotopes
are listed in Tables \ref{tab.hfalkaline}, \ref{tab.hflanthanide}.
Relatively few experiments have reported on the hyperfine structure of
these complex
atoms\cite{RN1,RN2,EH1,MA1,BMT,NBB,JOM}.

\begin{table}[!t]
\centering
\begin{tabular}{|l|c|c|}
\hline
isotope& nuclear spin & Rydberg spectroscopy experiments\\
\hline
$^{9}$Be & $3/2$ & \cite{RKM,YMT}\\
$^{25}$Mg & $5/2$ & \cite{BS1,WLJ}\\
$^{43}$Ca & $7/2$ & \cite{BLS,GHK} \\
$^{87}$Sr & $9/2$ & \cite{CGE,BMT,MMJ} \\
$^{135}$Ba, $^{137}$Ba & $3/2$& \cite{RN1,RN2,EH1,MA1} \\
\hline
\end{tabular}
\caption{
\label{tab.hfalkaline} Stable alkaline earth isotopes with hyperfine structure.
There have been a large number of measurements of the Rydberg spectra of
alkaline earth atoms. The cited references are representative and not intended to
be complete.  }
\end{table}

The Rydberg structure of multielectron atoms with hyperfine split
cores  has been only partially dealt with in earlier theoretical
works\cite{MA2,MA1,SL1,JQS,AGL,SLB}. These studies focused on obtaining
the energies for bound and autoionizing states as well as the dipole transition
operator from the ground or low-lying excited states.
The objective of this paper is to develop a detailed formalism that  can be
applied for calculation of single atom Rydberg properties and two-atom interaction
strengths. In addition to the energies and dipole transition operators, we
provide a framework to calculate the Zeeman shift and coupling as well as
the Stark shift and coupling between Rydberg states for atoms in weak
magnetic and electric fields. In addition to one atom properties, we
describe a formalism for obtaining the $C_n$ coefficients for the
Rydberg-Rydberg interactions and provide specific formulas for the
$C_5$ and $C_6$ matrices. The calculation of these interaction
matrices has not been discussed for Rydberg states attached to
hyperfine split thresholds.

The rest of the paper is organized as follows. Section~\ref{SecOneAtom}
contains the basic ideas for the frame transformations which form the
foundation for the rest of the developments. The frame transformation
for hyperfine split core states has been developed and has been applied
several times in experiments\cite{SL1,JQS,SLB,WHM,MSM,SRM}.
The inclusion of the frame transformation is for completeness and to
specify the notation used in the rest of this paper. Section~\ref{SecOneAtom}
goes beyond the previous developments by giving the expressions for
the Zeeman and Stark effect for Rydberg states attached to hyperfine
split core states. Section~\ref{SecTwoAtom} includes the derivation for
Rydberg-Rydberg interactions with specific derivation of the $C_5$ and
$C_6$ matrices. Section~\ref{SecHoEx} gives the parameters that are
known for $^{165}$Ho which will be used in example calculations.
Sections~\ref{SecHoExOne} and \ref{SecHoExTwo} give example
results for the one atom and two atom interactions for $^{165}$Ho.
Because the Rydberg properties of Ho are only partially known, these
calculations are presented more to demonstrate how to use the theory
than for the specific results. This is followed by a short summary.

\begin{table}[!t]
\centering
\begin{tabular}{|l|c|c|}
\hline
isotope& nuclear spin & Rydberg spectroscopy experiments\\
\hline
$^{159}$La & $7/2$ & \cite{XXH,SXX}\\
$^{141}$Pr& $5/2$ & \\
$^{143}$Nd, $^{145}$Nd& $7/2$ & \\
$^{149}$Sm& $7/2$ & \cite{JRB,WCY,ZDY,SSP}\\
$^{151}$Eu, $^{153}$Eu& $5/2$ & \cite{NRC,XDL,BRC,WSD} \\
$^{155}$Gd, $^{157}$Gd &$3/2$  & \cite{MOW, NBB,JOM,AD1} \\
$^{159}$Tb & $3/2$ & \\
$^{161}$Dy, $^{163}$Dy & $5/2$ &  \cite{XZH,SDN}\\ 
$^{165}$Ho & $7/2$ & \cite{HPL}\\
$^{167}$Er & $7/2$ & \\
$^{169}$Tm & $1/2$ & \cite{VAK}\\
$^{171}$Yb, $^{173}$Yb & $1/2,~ 5/2$ & \cite{CDM,ACD,AYN,ZDW}\\
$^{175}$Lu& $7/2$ & \cite{OK1,DZX,LLZ}\\
\hline
\end{tabular}
\caption{
\label{tab.hflanthanide} Stable or observationally stable  lanthanide isotopes
with hyperfine structure. Measurements of the ionization potentials for all the
lanthanides can be found in\cite{WSP}. The cited references are representative
and not intended to be complete.}
\end{table}

Atomic units are used unless explicitly stated otherwise.

\section{One atom theory}\label{SecOneAtom}

To organize the thinking about this type of system, we will first consider how
to describe the single atom Rydberg series. Also, most of the expressions for
matrix elements for Rydberg-Rydberg systems are present in the one atom
theory. One of the difficulties is to keep
track of all of the quantum numbers. In this section, we will use the symbol
$\alpha_c$ to indicate all of the quantum numbers in the core state except
the hyperfine angular momentum $F_c$. The Rydberg electron quantum
numbers will be denoted $n,s,\ell ,j$ for the principal quantum
number, spin, orbital angular momentum and total angular momentum.

One can get an idea of the principal quantum number where perturbers
attached to different hyperfine levels start to perturb each other by
setting the spacing of Rydberg energies, $1/\nu^3$, equal to
the spacing of the hyperfine energies $\Delta E_c$. The parameter
$\nu = n-\mu$ where $\mu$ is the quantum defect and $n$
is the principal quantum number. For a 10~GHz spacing,
$\Delta E_c = 1.52\times 10^{-6}$~atomic units (a.u.) which corresponds to
$\nu\simeq 87$. In the rare earth positive ions, the splitting of the
ionization thresholds can be larger than this which means the interactions
between channels can be at smaller $\nu$. As an example,
taking the Ho$^+$ $F_c=23/2$ state as $E=0$, the $21/2$ state
is at $-$17.6~GHz, the 19/2 state is at $-$34.0~GHz, the
17/2 state is at $-$48.9~GHz, etc.\cite{LSC}

\subsection{Bound states}

In this subsection, we will review the idea for how to find the bound
states and normalize them\cite{AGL}. For a Rydberg electron
attached to channel $|\Phi_i\rangle$, the coupled, real functions
$|\psi_{i}\rangle$ with unphysical boundary conditions at large $r$
can be written as
\begin{equation}
|\psi_{i}\rangle=\sum_{i'}|\Phi_{i'}\rangle (f_{i'}(r)\delta_{i',i} - g_{i'}(r) K_{i',i})
\end{equation}
where the $f,g$ are the energy normalized, radial Coulomb functions
which are regular,irregular at the origin
and ${\bf K}$ is the real, symmetric K-matrix. The Coulomb functions
$f_{i'},g_{i'}$ are for an orbital angular momentum $\ell_{i'}$ and
at energy $E - E_{c,i'}$ with $E$ being the total energy and
$E_{c,i'}$ is the energy of the core state $|\Phi_{i'}\rangle $ which encompasses
all of the degrees of freedom except the radial coordinate of the Rydberg
electron. The $|\psi_i\rangle $ function is
unphysical because the $f_{i'},g_{i'}$ functions diverge at large $r$ for closed channels
defined by $E<E_{c,i'}$.
It is understood
that this form only holds when the Rydberg electron is at distances,
$r$, larger than the radial size of the core state. The K-matrix
parameterizes the coupling between the $N$ different channels.
At large distances,
the $f,g$ vary rapidly with energy while the K-matrix will be
assumed to not vary rapidly with energy. When the total
energy is less than the threshold energy for all channels, the
$N$ different $|\psi_i\rangle$ can {\it not} be superposed in
a way to remove the exponential divergence at large $r$
for the Rydberg electron unless the total energy is at one of the
bound state energies. Both of the $f,g$ functions diverge
as $r\to\infty$ with
\begin{equation}
\lim_{r\to\infty} f_{i'}(r)/g_{i'}(r) \to -\tan (\beta_{i'})
\end{equation}
with $\beta _{i'} =\pi (\nu_{i'} -\ell_{i'})$ and the effective quantum
number defined by $E = E_{c,i'} - 1/[2\nu_{i'}^2]$ with $E$ being
the total energy and $E_{c,i'}$ is the energy of the $i'$-th core state.

At a bound state energy, $E_b$, the $|\psi_{i}\rangle$ can be superposed
so that the exponentially diverging parts of the wave function all 
cancel perfectly. This only leaves radial functions that exponentially
converge to 0 as $r\to\infty$. The superposition coefficients are
defined as
\begin{equation}
|\psi_b\rangle = \sum_{i} |\psi_{i}\rangle \frac{\cos (\beta_{i})}{\nu_{i}^{3/2}}A_{i,b}
\end{equation}
and the condition that determines them is\cite{AGL}
\begin{equation}\label{eqAdef}
\sum_{i} (\tan (\beta_{i'})\delta_{i',i}+K_{i',i}) \frac{\cos (\beta_{i})}{\nu_{i}^{3/2}}A_{i,b} = 0.
\end{equation}
This condition can be satisfied only for the energies where the determinant
of the term in parenthesis, $\tan\beta + K$, is zero. When this condition
is satisfied, the bound state function can be written as
\begin{equation}\label{eqPsiBnd}
|\psi_b\rangle = \sum_i |\Phi_i\rangle P_{\nu_i\ell_{i}}(r)A_{i,b}
\end{equation}
where the radial Coulomb function that goes to 0 at infinity is
\begin{equation}\label{eqCoulP}
P_{\nu_i\ell_{i}}(r)=(f_i\cos (\beta_i)+
g_i\sin (\beta_i))/\nu_i^{3/2}.
\end{equation}
In the limit that the K-matrix is slowly varying as a function of principal
quantum number (typically true for weakly bound Rydberg states),
the normalization condition is
\begin{equation}
\sum_i A^2_{i,b} = 1
\end{equation}
which, with Eq.~(\ref{eqAdef}), defines the $A_{i,b}$ within an
irrelevant, overall sign.

\begin{figure*}
\resizebox{160mm}{!}{\includegraphics{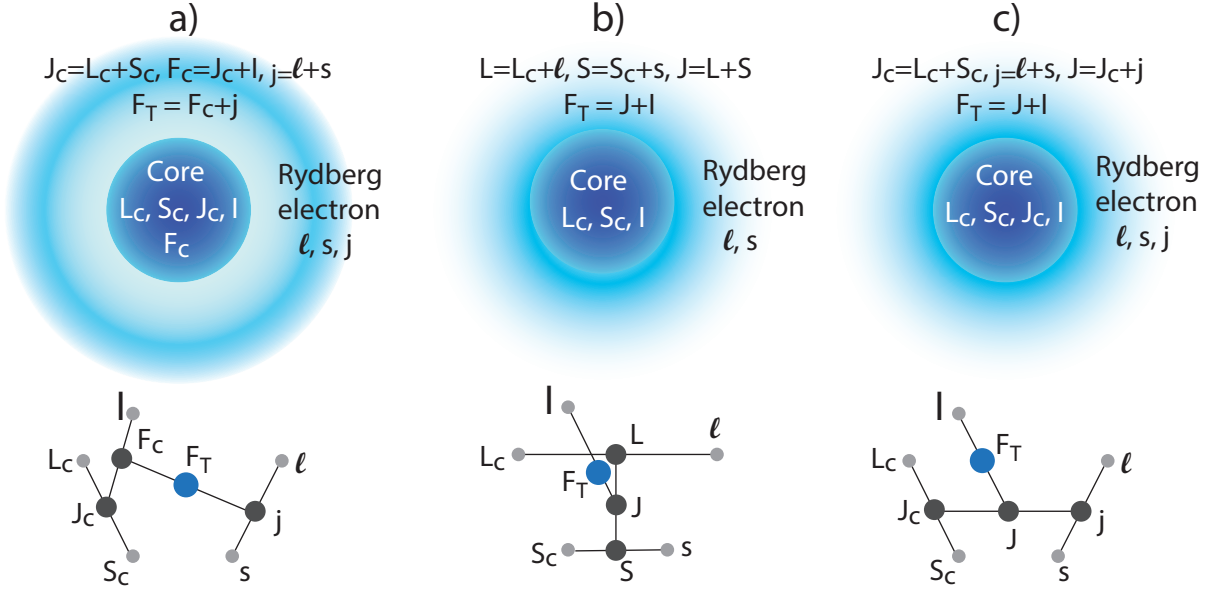}}
\caption{\label{FigFT}
A schematic showing the different angular momentum coupling schemes
for the frame transformations. The coupling in a) is appropriate when
the Rydberg electron is in the far outer region, $|{\rm out}\rangle$.
The coupling in b) is appropriate when all of the electrons are in the
core region, $|{\rm in}^{(LS)}\rangle$. The coupling in c) is appropriate
when the Rydberg is just outside of the core region but is not at
sufficient distance for the hyperfine interaction to give important
phase differences in the Coulomb functions, $|{\rm in}^{(jj)'}\rangle$.
}
\end{figure*}
\subsection{Frame transformation}\label{SecFT}

Since the K-matrix is real and symmetric, there are $N(N-1)/2$
parameters for the coupling of the channels. Unfortunately, it
is well beyond current computational resources to calculate the K-matrix with
sufficient accuracy to be useful for the lanthanides, for example. The experimental
energies for the bound states can be measured very accurately
and constrain the values of the K-matrix.
Because we assume the K-matrix has little variation with
energy, the many different energies for the bound states 
can lead to the case of having more energies than unknown
values in ${\bf K}$.
However, for complex atoms, the number of different free parameters
can be large which makes  the fitting process itself a difficult
numerical problem.

One solution is to use frame transformation ideas to limit the number
of free parameters to $\sim N$.\cite{AGL} The frame transformation
for a hyperfine split core state was developed in Ref.~\cite{JQS}
and applied to the hyperfine states of odd isotopes of Sr\cite{SLB}
or the hyperfine states of the heavier noble gas atoms.\cite{WHM,MSM,SRM}
There are several different ways for coupling the angular momentum
to obtain a frame transformation. The derivation in this section
describes the method we used.

The angular momentum coupling we used when the Rydberg electron is in the outer region,
Fig.~\ref{FigFT}a),
couples the total spin, $S_c$, and orbital angular momentum, $L_c$, of the core to
give a total angular momentum of the core, $J_c$, which is coupled
to the nuclear spin, $I$, to give the hyperfine angular momentum of
the core, $F_c$. The Rydberg electron has its spin, $s$, and orbital
angular momentum, $\ell$, coupled together to give a total angular
momentum, $j$. The total angular momentum of the core is coupled
to the total angular momentum of the Rydberg electron to give the
total angular momentum, $F_T$. We will symbolically write this coupling
scheme as
\begin{equation}\label{eqOut}
|{\rm out}\rangle = |(((S_cL_c)J_cI)F_c(s\ell )j)F_T\rangle
\end{equation}
with the ordering of the parenthesis indicating which angular momenta
are being coupled at each stage.

When the Rydberg electron is at short distances, then the LS-coupling
is more appropriate, Fig.~\ref{FigFT}b). Within this scheme, the total spin of the core
is coupled to the spin of the Rydberg electron to give the total spin, $S$,
the total orbital angular momentum of the core is coupled to the orbital
angular momentum of the Rydberg electron to give the total orbital
angular momentum, $L$. The total spin and total orbital angular momentum
are coupled to give the total angular momentum of all electrons, $J$,
which is then coupled to the spin of the nucleus to give the total
angular momentum, $F_T$. We will symbolically write this coupling
scheme as
\begin{equation}\label{eqInLS}
|{\rm in}^{(LS)}\rangle = |(((S_cs)S(L_c\ell )L)JI)F_T\rangle .
\end{equation}
Typically, the frame transformation would obtain the K-matrix in the
$|{\rm out}\rangle$ states by projecting onto the $|{\rm in}^{(LS)}\rangle$ states. However,
it seems likely that higher accuracy will be needed. So we will adopt
a method where there will be an intermediate jj-coupling, Fig.~\ref{FigFT}c). This coupling
will be to couple the spin of the core electrons to the orbital angular
momentum of the core electrons to obtain the total electronic angular
momentum of the core, $J_c$, and similarly for the Rydberg electron
giving its total angular momentum, $j$. These two angular momenta
are coupled together to give the total angular momentum of the
electrons, $J$, which is coupled to the spin of the nucleus to give
the total angular momentum of the atom. This will be represented as
\begin{equation}\label{eqInLSjj}
|{\rm in}^{(jj)}\rangle = |(((S_cL_c)J_c(s\ell )j)JI)F_T\rangle .
\end{equation}
By using this coupling, we can use the LS-coupled K-matrix to obtain a
jj-coupled K-matrix using a frame transformation. The advantage of
this intermediate step is that the resulting jj-coupled K-matrix can be
corrected as discussed in Sec.~\ref{SecCorr}.

The unitary matrix that arises from this step of the frame transformation is
\begin{eqnarray}\label{eqLS2jj}
\langle {\rm in}^{(LS)}|{\rm in}^{(jj)}\rangle&=& \langle ((S_cs)S(L_c\ell )L)J|((S_cL_c)J_c(s\ell )j)J\rangle\nonumber\\
&=&[S,L,J_cj]
\begin{Bmatrix}
S_c & s & S\\
L_c & \ell & L\\
J_c & j & J
\end{Bmatrix}
\end{eqnarray}
which is from Eq.~(6.4.2) of Ref.~\cite{ARE}.
The notation $[a,b,...]=\sqrt{(2a+1)(2b+1)...}$ and the $I$ and $F_T$ quantum
numbers drop out because they are in the same spot in the bra and in the ket.

The frame transformation approximation assumes that the channel coupling
between different $|{\rm in}^{(LS)}\rangle $ channels is small because LS-coupling
dominates the interaction. In this case the, K-matrix in jj-coupling is approximated as
\begin{equation}\label{eqKLS2jj}
K^{(jj)}_{aa'}=\sum_{LS} \langle {\rm in}^{(jj)}_a|{\rm in}^{(LS)}\rangle K^{(LS)}\langle {\rm in}^{(LS)}|{\rm in}^{(jj)}_{a'}\rangle
\end{equation}
where $K^{(LS)}=\tan [\pi\mu (LS) ]$.  For this expression,
there is a sum over all LS-coupled channels
that satisfy the angular momentum relations. The parameters $a,a'$ are
indicating the different jj-coupled channels

The jj-coupled K-matrices are then frame transformed to the $|{\rm out}\rangle$ channels
using the projection matrix
\begin{eqnarray}
\langle{\rm in}^{(jj)}|{\rm out}\rangle =\langle ((J_cj)JI)F_T|((J_cI)F_cj)F_T\rangle \nonumber\\
=(-1)^{3j+F_c+2J_c+I-J}[J,F_c]
\begin{Bmatrix}
j & J_c & J\\
I & F_T & F_c
\end{Bmatrix}
\end{eqnarray}
where the last step used Eq.~(6.1.5) of Ref.~\cite{ARE} and the earlier
steps use $\langle j_1m_1,j_2m_2|j_{12}m_{12}\rangle =(-1)^{j_1+j_2-j_{12}}
\langle j_2m_2,j_1m_1|j_{12}m_{12}\rangle $.
We have checked that the resulting expression for the composite
$\langle {\rm in}|{\rm out}\rangle$ gives a unitary matrix.
The K-matrix in the channels $|{\rm out}\rangle$ is obtained by a frame
transformation
\begin{equation}\label{eqKjj2hf}
K_{i',i}=\sum_{aa'}\langle {\rm out}_{i'}|{\rm in}^{(jj)}_a\rangle K^{(jj)}_{aa'}\langle {\rm in}^{(jj)}_{a'}|{\rm out}_i\rangle
\end{equation}
where the sum over all of the jj-coupled channels is indicated
by the $a,a'$.

\subsubsection{Corrections to the K-matrix}\label{SecCorr}

At the LS-coupling level, the quantum defects do not depend on the
$J$ or $F_T$ quantum numbers in the $|{\rm in}\rangle$ state. Thus, we
expect the $K^{(LS)}$ to depend only on the $L_c,S_c,\ell, S, L$. This
drastically reduces the number of free parameters. A test of the
accuracy of the approximation is in how well the spectra can be fit
with those parameters. While this should account for most of the
K-matrix, there are interactions in the heavier atoms that are not
encompassed by this approximation. The $K^{(jj)}$ can not be
exactly reproduced using Eq.~(\ref{eqKLS2jj}). One possible method
for improving the accuracy of the final K-matrix is to fit the levels
with an LS-to-jj frame transformation. Once a somewhat accurate
$K^{(jj)}$ is obtained, we can add a small correction to it to
improve the fit of the energy levels. Having an accurate $K^{(jj)}$
should be sufficient for most purposes because the hyperfine splittings
are so small that there should be almost no effect on the short
range K-matrix from dropping the hyperfine interaction.

\subsection{General one atom matrix elements}

For one atom matrix elements, there are two common situations
worth treating generally. The first is when the operator
$\hat{\zeta}$ only acts on the channel function $|\Phi_i\rangle$
{\it and} does
not change the $\ell $ of the Rydberg electron. For
example, when the atom is in a weak magnetic field, the
states with different $\ell$ are
not mixed. Even with these conditions, the radial integral for the
Rydberg electron is not trivial because it can involve different
binding energies. Such an integral was discussed in Ref.~\cite{BCC}.
Using the expression Eq.~(4.1.2) of Ref.~\cite{AGL} leads to
an expression for $\zeta_{bb'}=\langle\psi_b|\hat{\zeta}|\psi_{b'}\rangle$:
\begin{equation}\label{eqZeta}
\zeta_{b,b'}
=\sum_{i,i'}(A^T)_{b,i}\langle\Phi_i|\hat{\zeta}|\Phi_{i'}\rangle
O_{ib,i'b'}A_{i',b'}
\end{equation}
where the radial integral overlap integral gives
\begin{equation}
O_{ib,i'b'}=
\frac{2\sqrt{\nu_{ib}\nu_{i'b'}}}{\nu_{ib}+\nu_{i'b'}}
\frac{\sin  (\beta_{ib}-\beta_{i'b'})}{(\beta_{ib}-\beta_{i'b'})}.
\end{equation}
The subscript $b$ or $b'$ is added to the parameters
in the overlap because the functions
could be evaluated at a different energy if $E_b\neq E_{b'}$.
In the limit $\nu_{ib}=\nu_{i'b'}$, the overlap is simply 1. Because
the $E_b$ will typically have similar quantum defects, the overlap
will typically be small unless $\nu_{i,b}-\nu_{i',b'}\sim 0$.
To evaluate the matrix elements, the only new information needed
is the matrix elements of $\hat{\zeta}$ in the channel functions
$|\Phi_i\rangle$.

The other common situation is when the operator only acts on
the Rydberg electron. The specific case of most interest is when
the operator has the form $\hat{Q}^{(kq)}=r^kY_{k,q}(\Omega )$.
This operator has a contribution of size $\sim n^{2k}$ when it acts
on the Rydberg electron and of size $\sim 1$ when it acts on the
core electrons. Since the core contribution is relatively tiny, we will
only account for the contribution from the Rydberg electron. The matrix element
has the form
\begin{equation}\label{eqQcoup}
Q^{(kq)}_{b,b'}=\sum_{i,i'}(A^T)_{b,i}\langle\Phi_i|Y_{kq}|\Phi_{i'}\rangle
R^{(k)}_{ib,i'b'}A_{i',b}
\end{equation}
where the radial integral is
\begin{equation}
R^{(k)}_{ib,i'b'}=\int dr r^kP_{\nu_{i,b}\ell_{i}}(r)P_{\nu_{i',b'}\ell_{i'}}(r)
\end{equation}
with the radial functions defined in Eq.~(\ref{eqCoulP}). The upper
limit of integration is infinity. The lower limit is not 0 because the
form of the wavefunction in Eq.~(\ref{eqPsiBnd}) only holds when
$r$ is larger than the radial size of the core state. Since only a tiny
fraction of the radial integral accrues in this region, setting the lower
limit to the region larger than the small $r$ turning point of the
effective potential leads to sufficiently accurate results. The angular
integration is obtained analytically using the $|{\rm out}\rangle$ coupling
\begin{equation}
\langle\Phi_i|Y_{kq}|\Phi_{i'}\rangle =\delta_{c_i,c_{i'}}
\langle (F_c(s\ell )j)F_TM_T|Y_{kq}| (F_c(s\ell' )j')F'_TM'_T\rangle
\end{equation}
where the $\delta_{c_i,c_{i'}}$ means all of the core quantum numbers
are the same for $i$ and $i'$ and the rest of the matrix element can be
evaluated using Eqs.~(5.4.1), (5.4.5), and (7.1.8) of Ref.~\cite{ARE}
to obtain
\begin{eqnarray}
\langle\Phi_i|Y_{kq}|\Phi_{i'}\rangle & =&(-1)^\Lambda\delta_{c_i,c_{i'}}
\frac{[F_T,F'_T,j,j',\ell,\ell',k]}{\sqrt{4\pi}}\nonumber\\
&\null &\times
\begin{pmatrix}
F_T&k&F'_T\\
-M_T&q&M'_T
\end{pmatrix}
\begin{pmatrix}
\ell & k &\ell'\\
0&0&0
\end{pmatrix}\nonumber\\
&\null &\times
\begin{Bmatrix}
j&F_T&F_c\\
F'_T&j'&k
\end{Bmatrix}
\begin{Bmatrix}
\ell & j & s\\
j'&\ell'&k
\end{Bmatrix}
\end{eqnarray}
with $\Lambda = 2F_T-M_T+F_c+j+j'+\ell'+\ell+s$.
The first two three-$j$ symbols restrict $|F_T-F'_T|\leq k$
and $|\ell-\ell'|\leq k$ respectively. The second three-$j$ symbol
also restricts the sum $\ell+\ell'+k$ to be an even integer.
The last six-$j$ symbol restricts $|j-j'|\leq k$.

\subsection{Zeeman shifts and coupling}

It is often useful to add a magnetic field during an experiment to be
able to address only one state of a degenerate level. Thus, it is
worthwhile to obtain the Zeeman shifts and/or coupling between
states. In the section below, we will treat the case of two interacting
atoms. There, it's convenient to have the interatomic axis 
be defined as the $z$-direction. So in this section, we will allow the
magnetic field to be in an arbitrary direction. This case corresponds
to that covered by Eq.~(\ref{eqZeta}). The Zeeman Hamiltonian
can be written as
\begin{equation}
H_Z = \{\mu_B [\vec{L}_c + \vec{\ell}+g_s(\vec{S}_c+\vec{s})]-\mu_I\vec{I}\}\cdot \vec{B}
\equiv -\vec{\mu }\cdot\vec{B}
\end{equation}
where $\mu_B$ is the Bohr magneton, $g_s=2.002319...$, $\mu_I$ is
the nuclear magnetic moment.

Since the Zeeman Hamiltonian is a dot product of two vectors, we can
use the definition of tensor operator, Eq.~(5.1.3) of Ref.~\cite{ARE},
$V_{\pm 1} =\mp (V_x \pm i V_y)/\sqrt{2}$ and $V_0=V_z$ to obtain
\begin{eqnarray}
\langle\Phi_i |H_Z|\Phi_{i'}\rangle&=&\sum_{q=-1}^1B_q^*\langle\Phi_i |\mu^{(1)}_q|\Phi_{i'}\rangle\nonumber\\
&=& (B^{F_T,F'_T}_{M_T,M'_T})^*\langle\Phi_i ||\mu^{(1)}_q||\Phi_{i'}\rangle
\label{eqZeem1}
\end{eqnarray}
where
\begin{equation}
B^{F_T,F'_T}_{M_T,M'_T}=
 (-1)^{F_T-M_T}\sum_{q=-1}^1
\begin{pmatrix}
F_T & 1 & F_T'\\
-M_T & q &M_T'
\end{pmatrix}
B_q
\end{equation}
uses Eq.~(5.4.1) of Ref.~\cite{ARE}.

The $|\Phi_i\rangle$ have the angular momentum coupling of Eq.~(\ref{eqOut})
while the $\vec{\mu}$ is composed of operators acting on the $\ell ,s,L_c,S_c,I$
parts of $|\Phi_i\rangle$. The contribution of each of these terms to the matrix
element needs to be found separately. However, many of the operators involve
nearly the same steps as the others. Thus, many of the terms have the same
coefficients. The formulas below only use Eqs.~(5.4.3), (7.1.7) and (7.1.8)
of Ref.~\cite{ARE}. None of the angular momentum operators can change
the $\ell ,s,L_c,S_c,I$ which means all of the matrix elements are multiplied
by the quantity $\delta^{(5)}=\delta_{\ell \ell'}\delta_{ss'}\delta_{L_cL_c'}
\delta_{S_cS_c'}\delta_{II'}$. The reduced matrix elements are
\begin{eqnarray}
\langle\Phi_i||\ell^{(1)}||\Phi_{i'}\rangle &=&\delta^{(5)}{\cal G}_1{\cal G}_2\Lambda (\ell )\nonumber\\
\langle\Phi_i||s^{(1)}||\Phi_{i'}\rangle &=&\delta^{(5)}{\cal G}_1{\cal G}_3\Lambda (s )\nonumber\\
\langle\Phi_i||I^{(1)}||\Phi_{i'}\rangle &=&\delta^{(5)}{\cal G}_4{\cal G}_5\Lambda (I )\nonumber\\
\langle\Phi_i||L_c^{(1)}||\Phi_{i'}\rangle &=&\delta^{(5)}{\cal G}_4{\cal G}_6{\cal G}_7\Lambda (L_c )\nonumber\\
\langle\Phi_i||S_c^{(1)}||\Phi_{i'}\rangle &=&\delta^{(5)}{\cal G}_4{\cal G}_6{\cal G}_8\Lambda (S_c )
\label{eqZeem2}
\end{eqnarray}
where
\begin{eqnarray}
\Lambda (x) &=&\sqrt{(2x+1)(x+1)x}\nonumber\\
{\cal G}_1 &=&\delta_{J_cJ_c'}\delta_{F_cF_c'}(-1)^{F_c+j'+F_T+1}[F_T,F_T']
\begin{Bmatrix}
j&F_T&F_c\\
F_T'&j'&1
\end{Bmatrix}\nonumber\\
{\cal G}_2 &=&(-1)^{s+\ell+j+1}[j,j']
\begin{Bmatrix}
\ell & j&s\\
j' & \ell & 1
\end{Bmatrix}\nonumber\\
{\cal G}_3 &=&(-1)^{s+\ell +j'+1}[j,j']
\begin{Bmatrix}
s & j&\ell\\
j' & s & 1
\end{Bmatrix}\nonumber\\
{\cal G}_4 &=&\delta_{jj'}(-1)^{F_c+j+F_T'+1}[F_T,F_T']
\begin{Bmatrix}
F_c & F_T & j\\
F_T' & F_c' &1
\end{Bmatrix}
\nonumber\\
{\cal G}_5 &=&\delta_{J_cJ_c'}(-1)^{J_c+I+F_c+1}[F_c,F_c']
\begin{Bmatrix}
I & F_c & J_c\\
F_c' & I & 1
\end{Bmatrix}
\nonumber\\
{\cal G}_6 &=&(-1)^{J_c+I+F_c'+1}[F_c,F_c']
\begin{Bmatrix}
J_c & F_c & I\\
F_c' &J_c' & 1
\end{Bmatrix}
\nonumber\\
{\cal G}_7 &=&(-1)^{S_c+L_c+J_c+1}[J_c,J_c']
\begin{Bmatrix}
L_c & J_c & S_c\\
J_c' &L_c & 1
\end{Bmatrix}
\nonumber\\
{\cal G}_8 &=&(-1)^{S_c+L_c+J_c'+1}[J_c,J_c']
\begin{Bmatrix}
S_c & J_c & L_c\\
J_c' &S_c & 1
\end{Bmatrix}
\end{eqnarray}

\subsection{Electric field coupling}

An electric field can couple states of opposite parity whose angular
momenta differ by one or less. This situation corresponds to the case
covered by Eq.~(\ref{eqQcoup}). The electric field orientation will {\it not}
define the $z$-axis. Thus, we need to consider a general direction,
$\vec{\cal E}=({\cal E}_x,{\cal E}_y,{\cal E}_z)$. The Stark
Hamilgonian is $H_{st} = \vec{\cal E}\cdot\vec{r}$. To take
advantage of the angular momentum algebra, we will write this
Hamiltonian as
\begin{equation}
H_{st}=\sum_{q=-1}^1{\cal E}_q^*r_q
\end{equation}
where again we use $r_{\pm 1}=\mp (x\pm iy)/\sqrt{2}$ and $r_0=z$
as in Eq.~(5.1.3) of Ref.~\cite{ARE} and,
similarly, the ${\cal E}_{\pm 1}=\mp ({\cal E}_x\pm i {\cal E}_y)/\sqrt{2}$
and ${\cal E}_0={\cal E}_z$.
The matrix elements of the Stark Hamiltonian are
\begin{equation}
\langle\psi_b|H_{st}|\psi_{b'}\rangle=\sum_{q=-1}^1 {\cal E}_q^*Q^{(1q)}_{b,b'}
\end{equation}
using Eq.~(\ref{eqQcoup}).
If the electric field is taken to define the $z$-axis, the $M_T$ is a conserved
quantum number.

A common experimental situation is when the Rydberg atom experiences
a weak electric field. For states with substantial quantum defects, this
leads to weak coupling between states of opposite parity and quadratic
energy shifts. We will treat the possibility that two states, $b$ and $b'$,
of the same parity can be coupled through the mixing with opposite parity
states $b''$. This will only be relevant when the energy separation of
$b$ and $b'$ are small. The most common case occurs when
the electric field is not in the $z$-direction and the states are part of the
same degenerate, $F_T$, manifold. Using
second order perturbation theory, the weak electric field leads to nonzero
coupling between the states $b$ and $b'$:
\begin{equation}\label{eqH2Ef}
H^{(2)}_{b,b'}=\sum_{b''}\frac{\langle\psi_b|H_{st}|\psi_{b''}\rangle \langle\psi_{b''}|H_{st}|\psi_{b'}\rangle}{\bar{E}-E_{b''}}
\end{equation}
where $\bar{E}$ is the average energy of the two degenerate, or nearly degenerate,
states that are coupled
through the electric field: $\bar{E}=(E_b+E_{b'})/2$.
Diagonalizing the $H^{(2)}_{b,b'}$ gives the perturbative eigenstates in
the electric field. This quadratic energy shift with field strength is expressed
in the polarizability matrix.

\subsection{Dipole matrix elements to ``ground'' states}

The transition dipole matrix element that excites the atom from a compact
initial state into the Rydberg state is also nearly impossible to calculate
from first principles. It is possible to fit the transition matrix elements
using the oscillator strength to many different Rydberg states. However,
there will be $N$ different matrix elements which can lead to a very
complicated calculation to obtain the best values\cite{RDC}.

A way to reduce the number of parameters and/or find a decent starting
point for the fit is to use the coupling for the $|{\rm in}\rangle$ channels to obtain
approximate matrix elements. The unitary $\langle {\rm in}|{\rm out}\rangle$
frame transformation
can be used to obtain the matrix elements used for the oscillator strengths.
The different atoms and different initial states can lead to different recoupling
schemes. Thus, it is impossible to lay out a general formula for recoupling.
Instead, we will work through a recently measured case\cite{HPL}
as a demonstration for how this might be done.

\subsubsection{Holmium photo-excitation}\label{SecHoSig}

Reference~\cite{HPL} measured the Rydberg $s,d$ series in Ho starting
from the $J=17/2$ state 24,360.81 cm$^{-1}$ above the ground state.
The NIST data tables\cite{KRR}
gives the coupling as $4f^{11}(^4{\rm I}^o_{15/2})6s6p(^1P^o_1)$.
Because of the complicated electronic correlations, the accuracy of this
designation is uncertain. However, the designation of $J=17/2$
should be accurate. Thus, it seems that the main correlation will be mixing with
the three states $4f^{11}(^4{\rm I}^o_{15/2})6s6p(^3{\rm P}^o_{1,2})$ and
$4f^{11}(^4{\rm I}^o_{13/2})6s6p(^3{\rm P}^o_2)$. We will treat the dipole matrix
element as arising from the superposition of these four states with unknown
coefficients. We will denote any of these four states with the symbol
$|{\rm gr}\rangle$. The final states are $s,d$ states attached to the
 $4f^{11}(^4{\rm I}^o_{15/2})6s_{1/2}$ threshold with $J_c=8$. The
 nuclear spin $I=7/2$.

The dipole matrix element will be to the $|{\rm in}^{(LS)}\rangle$ states, Eq.~(\ref{eqInLS}).
However, the initial states track an extra electron over that for the
$|{\rm in}^{(LS)}\rangle$ states and the couplings are different. The basic idea is to recouple
the electrons in the initial state $|{\rm gr}\rangle$ to achieve the same type of coupling
as for $|{\rm in}^{(LS)}\rangle$. We then use the dipole operator, $D^{(1)}_q$, with the
angular coupling scheme to obtain the form of the matrix element.

The starting coupling scheme of the initial state is a partial spin of the core,
$\bar{S}_c$, coupled to a partial orbital angular momentum of the core,
$\bar{L}_c$, to give a partial total angular momentum of the core, $\bar{J}_c$.
For the Ho example, $\bar{S}_c=3/2$, $\bar{L}_c =6$, $\bar{J_c}=15/2$.
The other core electron spin, $s'_c$, is coupled to the spin of the outer
electron, $s$, to give $S_{co}$ (for Ho, these are $1/2$, $1/2$ and either 0 or 1).
The other core electron orbital angular momentum, $\ell'_c$, is coupled
to the orbital angular momentum of the outer electron, $\ell_i$, to give
$L_{co}$ (for Ho, these are 0, 1, and 1). The $S_{co}$ is coupled to
the $L_{co}$ to give a $J_{co}$ (for Ho, $J_{co}=1$ or 2). The $\bar{J}_c$
is coupled to the $J_{co}$ to give the total electronic angular momentum,
$J_i$ (for Ho, this is 17/2). This is then coupled to the nuclear spin, $I$,
to give the initial total angular momentum of the atom, $F_{Ti}$. This
can be represented as
\begin{equation}
|{\rm gr}\rangle=|(((\bar{S}_c\bar{L}_c)\bar{J}_c((s'_cs)S_{co}(\ell'_c\ell_i)L_{co})J_{co})J_iI)F_{Ti}\rangle
\end{equation}

The first step is to recouple in $|{\rm gr}\rangle$ the spins and orbital angular momentum to
get a total spin and total orbital angular momentum:
\begin{equation}
|{\rm gr}\rangle=\sum_{L_i}{\cal N}_1 |(((\bar{S}_cS_{co})S(\bar{L_c}L_{co})L_i)J_iI)F_{Ti}\rangle
\end{equation}
where
\begin{eqnarray}
{\cal N}_1&=&\langle ((\bar{S}_cS_{co})S(\bar{L_c}L_{co})L)J_i|((\bar{S}_c\bar{L}_c)\bar{J}_c(S_{co}L_{co})J_{co})J_i\rangle\nonumber\\
&=&[\bar{J}_c,J_{co},S,L_i]
\begin{Bmatrix}
\bar{S}_c &\bar{L}_c&\bar{J}_c\\
S_{co} & L_{co}& J_{co}\\
S & L_i & J_i
\end{Bmatrix}
\end{eqnarray}
is from Eq.~(6.4.2) of Ref.~\cite{ARE}. Actually, the sum should also be over $S_i$, but
we have used the fact that the dipole matrix element below will give a term with
$\delta_{S_i,S}$. The second step is to recouple the
spins from $(\bar{S}_c(s'_cs)S_{co})S$ to $((\bar{S}_cs'_c)S_cs)S$
which gives a six-$j$ coefficient
\begin{equation}
{\cal N}_2 = (-1)^{\bar{S}_c+s'_c+s+S}
\begin{Bmatrix}
\bar{S}_c & s'_c & S_c\\
s&S&S_{co}
\end{Bmatrix}
\end{equation}
from Eq.~(6.1.5) of Ref.~\cite{ARE}. A similar recoupling for the orbital
angular momentum gives
\begin{equation}
{\cal N}_3 = (-1)^{\bar{L}_c+\ell'_c+\ell_i+L_i}
\begin{Bmatrix}
\bar{L}_c & \ell'_c & L_c\\
\ell_i&L_i&L_{co}
\end{Bmatrix}
.
\end{equation}

The electrons in the $|{\rm gr}\rangle$ and the $|{\rm in}^{(LS)}\rangle$ are now in the same ordering which
allows the computation of the matrix element using standard angular momentum
recoupling. The dipole operator acts on the $\ell_i$, transitioning it to $\ell$.
\begin{eqnarray}\label{EqDip}
\langle {\rm gr}|&D&^{(1)}_q|{\rm in}^{(LS)}\rangle =\sum_{L_i}{\cal N}_1{\cal N}_2{\cal N}_3\langle F_{Ti}M_{Ti}|D^{(1)}_q|F_TM_T\rangle \nonumber\\
&=&\sum_{L_i}{\cal N}_1{\cal N}_2{\cal N}_3{\cal N}_4\langle (J_iI)F_{Ti}||D^{(1)}||(JI)F_T\rangle\nonumber\\
&=&\sum_{L_i}{\cal N}_1{\cal N}_2{\cal N}_3{\cal N}_4 {\cal N}_5 \langle (SL_i)J_i||D^{(1)}|(SL)J\rangle\nonumber\\
&=&\sum_{L_i}{\cal N}_1{\cal N}_2{\cal N}_3{\cal N}_4 {\cal N}_5 {\cal N}_6 \langle (L_c\ell_i)L_i||D^{(1)}||(L_c\ell )L\rangle\nonumber\\
&=&\sum_{L_i}{\cal N}_1{\cal N}_2{\cal N}_3{\cal N}_4 {\cal N}_5 {\cal N}_6 {\cal N}_7 \langle \ell_i||D^{(1)}||\ell \rangle
\end{eqnarray}
where at each step we have only shown the relevant angular momenta. The
coefficients are
\begin{equation}
{\cal N}_4 = (-1)^{F_{Ti}-M_{Ti}}
\begin{pmatrix}
F_{Ti}&1&F_T\\
-M_{Ti}&q&M_T
\end{pmatrix}
\end{equation}
from Eq.~(5.4.1) of Ref.~\cite{ARE},
\begin{equation}
{\cal N}_5=(-1)^{J_i+I+F_T+1}[F_{Ti},F_T]
\begin{Bmatrix}
J_i&F_{Ti}&I\\
F_T&J&1
\end{Bmatrix}
\end{equation}
from Eq.~(7.1.7) of Ref.~\cite{ARE},
\begin{equation}
{\cal N}_6=(-1)^{S+L+J_i+1}[J_i,J]
\begin{Bmatrix}
L_i& J_i&S\\
J & L&1
\end{Bmatrix}
\end{equation}
from Eq.~(7.1.8) of Ref.~\cite{ARE}, and
\begin{equation}
{\cal N}_7=(-1)^{L_c+\ell +L_i+1}[L_i,L]
\begin{Bmatrix}
\ell_i & L_i&L_c\\
L & \ell&1
\end{Bmatrix}
\end{equation}
from Eq.~(7.1.8) of Ref.~\cite{ARE}.

The only unknown coefficient is the last reduced matrix
element which will depend on which $\ell$ is excited for
the $|{\rm in}\rangle$ channel. To a good approximation, the $ \langle \ell_i||D^{(1)}||\ell \rangle$ 
is independent of the other angular momenta in the $|{\rm in}\rangle$ channel.
For the Ho example, there will be one reduced matrix element for
$\ell =0$ and a different one for $\ell =2$.

\begin{table}[!t]
\centering
\begin{tabular}{| c |  r  |  r  |  r  |}

\hline
\null\enskip\quad $S,\ell,L,J,F_T$\quad\enskip\null &\null\enskip 0,1\enskip\null &\null\enskip  1,1\enskip\null &\null\enskip  1,2\enskip\null  \\
\hline
3/2,0,6,15/2,11 &-2.28 &0.97 &-1.09\\
5/2,0,6,15/2,11 &0.00 &-1.86&0.35\\
5/2,0,6,15/2,11 &0.00 & 0.27&0.39\\
\hline
5/2,0,6,17/2,12 &0.00 &1.52 &2.21\\
\hline
3/2,2,7,17/2,12 &-1.13 &0.48 &-0.54 \\
3/2,2,8,17/2,12 &0.26 &-0.11 & 0.12\\
3/2,2,8,19/2,12 &-0.31 &-0.13 & -0.15\\
5/2,2,6,17/2,12 &0.00 &-1.22 & -1.24\\
5/2,2,7,17/2,12 &0.00 &-0.49 & 0.63\\
5/2,2,7,19/2,12 &0.00 &-0.22 & -0.21\\
5/2,2,8,17/2,12 &0.00 &0.28 & -0.15\\
5/2,2,8,19/2,12 &0.00 &-0.20 &0.11 \\
5/2,2,8,21/2,12 &0.00 &0.00 & 0.00\\
\hline
3/2,2,8,19/2,13 & -1.84 & 0.78&-0.88 \\
5/2,2,7,19/2,13 &0.00 &-1.29 & -1.22\\
5/2,2,8,19/2,13 &0.00 &-1.19 &0.63 \\
5/2,2,8,21/2,13 &0.00 &0.00 & 0.00\\
\hline
\end{tabular}
\caption{
\label{TabDip}
Example of coefficients for the dipole matrix element, Eq.~(\ref{EqDip}), assuming the
$\langle\ell_i||D^{(1)}||\ell\rangle = 1$ and without the ${\cal N}_4$
term. The three possible initial states have $F_{Ti}=12$ and
are $S_{co},J_{co}$ equaling
0,1 and 1,1 and 1,2. The different possible angular couplings  of $|{\rm in}^{(LS)}\rangle$
are denoted
by the angular momenta $S,\ell , L,J,F_T$. The horizontal lines denote the channels
with the same $\ell$ and $F_T$.}
\end{table}

Table~\ref{TabDip} gives the coefficients for the Ho example discussed above.
The values are a numerical calculation of Eq.~(\ref{EqDip}), assuming the
$\langle\ell_i||D^{(1)}||\ell\rangle = 1$ and without the ${\cal N}_4$
term. The ${\cal N}_4$ term is a simple 3-j factor and is the only term that
depends on $M_T$ and the polarization of the light.
The state $|{\rm gr}\rangle$ has the coupling $4f^{11}(^4I^o_{15/2})6s6p(^{2S_{co}+1}P_{J_{co}})$
with $J=17/2$, $I=7/2$, and $F_{Ti}=12$. There are three allowed cases:
$S_{co},J_{co}=(0,1)$, (1,1), and (1,2). These are the three different columns.
The rows correspond to the different $|{\rm in}^{(LS)}\rangle$ channels, Fig.~\ref{FigFT}b).
For this case, the channels are $(4f^{11}6s(^5I^o)n\ell)^{2S+1}L_J$ with $J$
coupled to $I=7/2$ to give $F_T$. For this case, the parameters can have the
values $S=3/2,5/2$, $\ell = 0,2$, $L=6,7,8$, $J=15/2,17/2,19/2,21/2$, and
$F_T = 11,12,13$. We did not include the 16 channels with $\ell = 2$ and $F_T=11$
for space reasons. There are no channels with $\ell = 0$ and $F_T=13$.
Most of the terms are between 0.1 and $\sim 1$ except
for a number that are identically zero due to angular momentum restrictions.
Note that the coupling for the $|{\rm gr}\rangle$ state in the NIST
data tables\cite{KRR} corresponds to the column 0,1 which has most of the
matrix elements exactly 0.

\section{Two Atom Theory}\label{SecTwoAtom}

This section is an extension of Ref.~\cite{VJP} which itself extended the
treatment of Rydberg-Rydberg interactions to the case of alkaline-earth
atoms with $I=0$. Unlike the alkali atoms which do not have a substantial angular
momentum for the core, the alkaline-earth atoms have an extra
core electron. This extra electron gives a spin-1/2 which the Rydberg
electron can couple to. This introduces extra terms in the matrix elements
which changes the Rydberg-Rydberg interactions.

Unlike the case for the $I=0$ alkaline-earth atoms, the extra core
electrons for the rare earth and odd isotope alkaline earth atoms
will give both hyperfine shifts and perturbed Rydberg series.
Thus, the expressions are somewhat more complicated. The derivation
below is in the most general form and is applicable to any atom. Most of the
examples discussed here have been for rare earth atoms. However,
the treatment below is also applicable to, for example, the strontium isotope
$^{87}$Sr which has $I=9/2$ and 7\% abundance; the Sr$^+$ has two
hyperfine states with $F_c=4,5$ with a splitting of $\sim 5$~GHz.

As with Ref.~\cite{VJP}, the largest error in the treatment below comes
from the lack of knowledge about the K-matrix. To the extent that the
K-matrix can be known as accurately as for the alkali or alkaline earth
atoms, then the resulting parameters (e.g. $C_n$ coefficients)
will be more accurate. However, for a given accuracy in the K-matrix,
the atom-atom interactions will tend to be less accurate compared to those
of the alkali
atoms simply due to the more complex Rydberg series, as will be shown below.

In the calculations below,
the atom-atom separation vector is assumed to lie along the $z$-axis.

\subsection{Two atom matrix elements}

Citing results in Refs.~\cite{MER,PRF,ADD}, Ref.~\cite{WTM} gives a multipole expansion
of the terms that couple Rydberg states in pairs of atoms in their Eqs.~(6-8).
In this expression is the product $p^{(1)}_{\kappa_1q}p^{(2)}_{\kappa_2 -q}$
where $p^{(i)}_{\kappa q}=r_i^{\kappa}Y_{\kappa q}(\Omega_i)\sqrt{4\pi /(2\kappa +1)}$. These matrix elements are
exactly the case treated in Eq.~(\ref{eqQcoup}) above. Supposing the
two atom state is written as $|b_1b_2\rangle$, the matrix element is
\begin{equation}
\langle b_1b_2|p^{(1)}_{\kappa_1q}p^{(2)}_{\kappa_2 -q}|b_1'b_2'\rangle
=\frac{4\pi}{[\kappa_1,\kappa_2]}Q^{(\kappa_1q)}_{b_1,b_1'}Q^{(\kappa_2-q)}_{b_2,b_2'}
\end{equation}
with the expressions for $Q$ given below Eq.~(\ref{eqQcoup}) and
$[\kappa_1,\kappa_2]$ defined below Eq.~(\ref{eqLS2jj}).

We give explicit expressions for the leading terms in the Rydberg-Rydberg
interaction at large separations, $R$, in the next two sections.

\subsubsection{$C_5$ coefficients}

The leading order Rydberg-Rydberg interaction at very large distances
leads to a coupling between states of the form: $C_{5;b_1b_2,b_1'b_2'}/R^5$.
In general, the $C_5$ coefficient is a matrix that couples the different
pair states. The $C_5$ is nonzero when the Rydberg orbital angular momentum
$\ell\geq 1$ and the total angular momentum is also $F_T\geq 1$.
Using Eq.~(7) of Ref.~\cite{WTM} with $\kappa_1=\kappa_2=2$ gives
\begin{equation}\label{eqC5}
C_{5;b_1b_2,b_1'b_2'}=\frac{4\pi}{5}\sum_{q=-2}^2\begin{pmatrix} 4\\ 2+q\end{pmatrix}
Q^{(2q)}_{b_1,b_1'}Q^{(2-q)}_{b_2,b_2'}
\end{equation}
where the binomial coefficient is 1 for $q=\pm 2$, 4 for $q=\pm 1$ and 6
for $q=0$.

\subsubsection{$C_6$ coefficients}

The $C_6$ coefficient arises from the perturbative interaction between the
atoms through the dipole-dipole term, $\kappa_1=\kappa_2=1$.
Using Eq.~(7) of Ref.~\cite{WTM} with $\kappa_1=\kappa_2=1$ gives
the matrix elements for the dipole-dipole interaction
\begin{equation}
\langle b_1b_2|V^{(dd)}|b_1'b_2'\rangle =\frac{C_{3;b_1b_2,b_1'b_2'}}{R^3}
\end{equation}
with
\begin{equation}\label{eqC3}
C_{3;b_1b_2,b_1'b_2'}=-\frac{4\pi}{3}\sum_{q=-1}^1\begin{pmatrix} 2\\ 1+q\end{pmatrix}
Q^{(1q)}_{b_1,b_1'}Q^{(1-q)}_{b_2,b_2'}
\end{equation}
where the binomial coefficient is 1 for $q=\pm 1$ and 2 for $q=0$.

The $C_6$ coefficient arises from the 2-nd order perturbative coupling
through pair Rydberg states of opposite parity. The coupling between
different pair states, $|b_1b_2\rangle$ and $|b''_1b''_2\rangle$,
only has a substantial effect when the initial and
final energies are nearly equal $E_{b_1}+E_{b_2}\simeq E_{b''_1}+E_{b''_2}$.
The coefficient is
\begin{equation}
C_{6;b_1b_2,b_1''b_2''}=\sum_{b_1'b_2'}\frac{C_{3;b_1b_2,b_1'b_2'}C_{3;b_1'b_2',b_1''b_2''}}{\bar{E}-E_{b_1'}-E_{b_2'}}
\end{equation}
where the $C_3$ coefficients are defined in Eq.~(\ref{eqC3}) and
$\bar{E}$ is the average energy of the two pair states:
$\bar{E} = (E_{b_1}+E_{b_2}+ E_{b''_1}+E_{b''_2})/2$.

As with the alkali and alkaline-earth atoms, the $C_6$ coefficient can be
strongly dependent on the pair states because there can be near degeneracies
in the energy denominator. For atoms with hyperfine split core states, there
are many more Rydberg states at each energy due to the additional multiplicity
from the core. This might lead to more states with large $C_6$ coefficients.
But it also points to the difficulty in the calculation, because even small changes
to the Rydberg energies or changes to the character of the Rydberg state
might strongly change the $C_6$.

Even when restricting the $C_6$ to the case where the two initial and
two final states are degenerate, $E_{b_1}=E_{b_2}= E_{b''_1}=E_{b''_2}$,
there can be a substantial number of states that couple through the
$C_6$. This is a somewhat more complicated version of Ref.~\cite{WS1}
for alkali atoms. We have not analytically analyzed the possible cases
as was done in Ref.~\cite{WS1}. The results shown below were obtained by
numerical calculation of the sum followed by a numerical diagonalization
of the states with $M_1+M_2$ the same.

\section{A rare earth example: $^{165}$Ho}\label{SecHoEx}

The treatment described above requires accurate atomic data to constrain the
parameters, K-matrix and dipole matrix elements, needed to calculate the
energies, oscillator strengths, $C_5$-coefficients, etc. Although this data does not
exist at this time, it is worthwhile to use the formulas above for a specific case
to provide an example of how they might be applied. In the calculations below,
we will not correct the K-matrix in Eq.~(\ref{eqKLS2jj}) but will directly use
that result to obtain the final K-matrix, Eq.~(\ref{eqKjj2hf}). Thus, the frame transformation
will proceed from the coupling in Fig.~\ref{FigFT}b) to that of \ref{FigFT}c); this
K-matrix will then utilize the frame transformation from the coupling in
Fig.~\ref{FigFT}c) to that of \ref{FigFT}a).

We will use Ho as an example. Although there has been a high precision study
of some of the $s$- and $d$-Rydberg series in Ref.~\cite{HPL}, most of the
Rydberg series are missing. Thus, the rough size of the quantum defects is
known but their dependence on $L,S,L_c,\ell ,$ etc is not constrained. This means
the size of the series interactions can not be accurately predicted.

However, the hyperfine splitting of the ionization thresholds
and the angular momenta of the
channels are well known. Since the number of states and the thresholds to which
they belong are perfectly constrained, the results below should be considered
a cartoon of the Rydberg state properties.

\subsection{Well known parameters}

From the NIST data tables,\cite{KRR} the Ho$^+$ ground state has the character
of $4f^{11}(^4{\rm I}^o_{15/2})6s_{1/2}$ $J_c=8$. They alternatively classify
the state as $(4f^{11}6s)^5{\rm I}^o_8$. Since the LS-to-jj frame transformation
needs all of the $^5{\rm I}^o$ core states, we also include the $J_c=7$ state
at 5617.04~cm$^{-1}$, the $J_c=6$ state at 5849.74~cm$^{-1}$,
the $J_c=5$ state at 8850.55~cm$^{-1}$, and the $J_c=4$ state
at 10838.85~cm$^{-1}$. Unlike the ground state, the smaller $J_c$ states
are not very pure in LS-coupling. But this is not important because the
states attached to these thresholds have small $\nu < 4.5$ which means
they do not contribute rapid energy dependence to the atomic parameters.
However, there is another state of Ho$^+$ that mainly has $^3{\rm I}^o$ symmetry and
$J_c=7$ at 637.40~cm$^{-1}$. A Rydberg state attached to this threshold
would have $\nu\simeq 13.1$ in the threshold region. Thus, there could
be perturbers attached to this threshold that would cause substantial
energy dependence to the quantum defects of the high Rydberg states.
In fact, there is a somewhat sharp perturber of an $s$-Rydberg series
near $n\sim 50$. We will discuss how to treat this type of perturber below.

The ground state threshold has hyperfine splitting from the nuclear spin
with $I=7/2$. Thus, the ground state core hyperfine angular momentum ranges from
$F_c=9/2$ to 23/2. The energies of the hyperfine states are at
\begin{equation}
E_{HF}=A\frac{K}{2}+B\frac{3K(K+1)-4I(I+1)J_c(J_c+1)}{8IJ_c(2I-1)(2J_c-1)}
\end{equation}
with $K=F_c(F_c+1)-J_c(J_c+1)-I(I+1)$ and
$A=52.61\times 10^{-3}$~cm$^{-1}$ and $B=-53.9\times 10^{-3}$~cm$^{-1}$
from Ref.~\cite{LSC}. In the calculations below, we will shift the hyperfine
energies by the energy of the $F_c=23/2$ state because
Ref.~\cite{HPL} reported their Rydberg energies relative to the
threshold with largest $F_c$. For the higher LS thresholds, we used the
same value of $A$ and $B$ because the hyperfine splitting is
irrelevant for the small $\nu$ states attached to those thresholds.

From the NIST tables, we now list the angular momentum quantum numbers
used in the calculations below. Because of the ion ground state, we use
$S_c=2$, $L_c=6$, $J_c=8,7,6,5,4$, $I=7/2$, and $s=1/2$ for the $|{\rm out}\rangle$
channels. The $\ell$ and $j$ depends on the Rydberg series being modeled.

As an example of the quantum numbers that can contribute, we examine
the photo-excitation case of Ref.~\cite{HPL}.
They measured the Rydberg $s,d$ series in Ho starting
from the $J_i=17/2$ initial state 24,360.81 cm$^{-1}$ above the ground state
in the highest hyperfine state of $F_{Ti}=12$. The dipole selection rules
means they can excite to $F_T=11,12,13$.
The NIST data tables gives the coupling as $4f^{11}(^4{\rm I}^o_{15/2})6s6p(^1{\rm P}^o_1)$
and they excited the $6p$ electron to the $s$- and $d$-Rydberg series.
Only examining the $|{\rm out}\rangle$ channels attached to the ground state of Ho$^+$
using the coupling of Fig.~\ref{FigFT}a),
we can list all of the channels with $s$-Rydberg series: for $F_T=13$ none,
for $F_T=12$ is 1 attached to the $F_c=23/2$, for $F_T=11$ is 1 attached to the
$F_c=23/2$ and 1 to the 21/2. Similarly for the $d$-Rydberg series using the notation
($F_c,j$):
for $F_T=13$ are (23/2,3/2), (23/2,5/2), (21/2,5/2), for $F_T=12$ are
(23/2,3/2), (23/2,5/2), (21/2,3/2), (21/2,5/2), (19/2,5/2), and for $F_T=11$
are (23/2,3/2), (23/2,5/2), (21/2,3/2), (21/2,5/2), (19/2,3/2), (19/2,5/2), (17/2,5/2).
Thus, for $F_T=13$, there are 3 Rydberg series (2 attached to 23/2 and
1 attached to 21/2); for $F_T=12$, there are 6 Rydberg series (3 attached to 23/2,
2 attached to 21/2, and 1 attached to 19/2); for $F_T=11$, there are
9 Rydberg series (3 attached to 23/2,
3 attached to 21/2, 2 attached to 19/2, and 1 attached to 17/2).

\subsection{Not well known parameters}

From Ref.~\cite{HPL}, an $s$-series quantum defect attached to
the $F_c=23/2$ threshold is $\mu_s\simeq 4.34$
but with a perturber near $n\sim 50$. From the previous section, there
are two $s$-series attached to this threshold, one with $F_T=12$ and one
with $F_T=11$. It is not clear which is the series with the perturber
but the following energy argument suggests the $F_T=11$ is the
perturbed series.
There is no information about the series attached to
the $F_c=21/2$ threshold but the quantum defect should be similar
to the series attached to the $F_c=23/2$: $\mu_s\sim 4.25-4.45$.
Also, we do not know if this series is perturbed by the same perturber
of the measured series or a different perturber. Interestingly, a perturber
attached to the threshold at 637.40~cm$^{-1}$ gives a $\nu =12.63=17-4.37=n-\mu$
which suggests an $s$-Rydberg state. Assuming the perturber is
attached to this $J_c=7$ threshold, an
$s_{1/2}$ Rydberg electron can at most give $J=15/2$; combined
with the $I=7/2$, the largest total angular momentum, $F_T$, could be
11. Thus, we would expect the $F_T=12$ series to be unperturbed but
both of the $F_T=11$ series to be perturbed.

Reference~\cite{HPL} measured several $d$-series quantum defects
attached to the $F_c=23/2$ threshold. These quantum defects range from
$\mu_d\simeq 2.7-2.82$. This will give a range of allowed $d$-state
quantum defects. Unfortunately, there is not much information about
the interactions between the Rydberg series so which of the quantum
defects are assigned which value is not known.

References~\cite{STM,FTD} provide crude estimates for $s,p,d,f$ quantum
defects for all atoms. The estimates for the $s,d$ quantum defects
are approximately those measured in Ref.~\cite{HPL}. Thus, we will use
their estimates of
quantum defects for the other angular momenta: $\sim$3.75 for the $p$- and $\sim$1.0 for the $f$-quantum
defects. Note that the $p$-quantum defects are nearly the same as
the $d$-quantum defects but shifted by an integer. If the $p$-quantum
defects are near this value, then the $d$ Rydberg series will have
very large polarizabilities and, perhaps, very large $C_6$ coefficients.

\section{One atom example results}\label{SecHoExOne}

\subsection{Ho energy levels}

\begin{figure}
\resizebox{80mm}{!}{\includegraphics{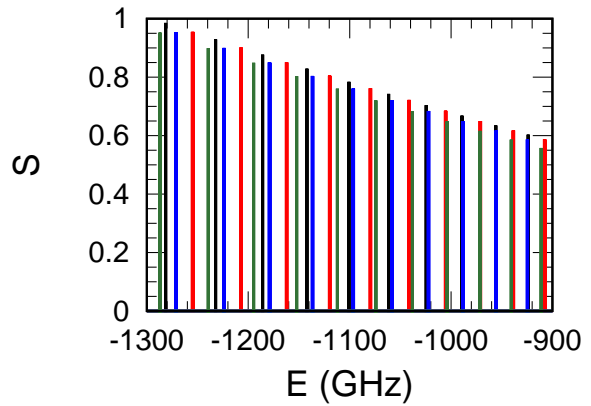}}
\caption{\label{FigEsimp1}
A plot of the uncorrelated energies of Ho $F_T=12$ $s$- and $d$-Rydberg
series. The energy is relative to the $F_c=23/2$ threshold which is the
highest hyperfine energy of the ground state.
The $S\propto 1/\nu^3$. The range -1300~GHz to -900~GHz corresponds
to $\nu = 50.3$ to $\nu = 60.5$. The purple line is the $s$-Rydberg series
attached to the $F_c=23/2$ threshold. The others are the two $d$-series
attached to the 23/2 threshold (purple), two series attached to the 21/2 threshold (blue),
and one series attached to the 19/2
threshold (green). There are several places showing the near degeneracy
of states attached to different thresholds which could lead to strong state
mixing.
}
\end{figure}

\begin{figure}
\resizebox{80mm}{!}{\includegraphics{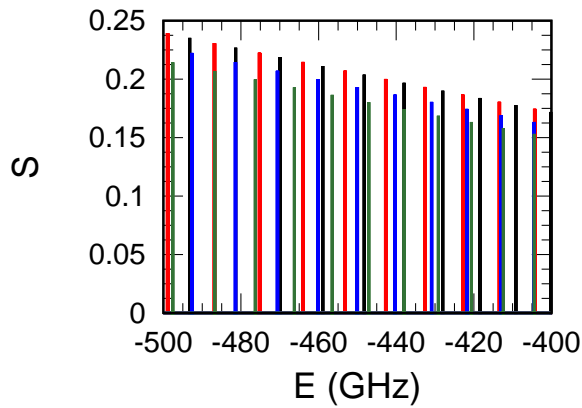}}
\caption{\label{FigEsimp2}
Same as Fig.~\ref{FigEsimp1} but for a smaller range.
The range -500~GHz to -400~GHz corresponds
to $\nu = 81.1$ to $\nu = 90.7$.
}
\end{figure}

In Figs.~\ref{FigEsimp1} and \ref{FigEsimp2} we show a simple stick
drawing of where uncorrelated Rydberg states appear in the spectrum
to give an idea of the complications possible. The plots are for
$F_T=12$ which has 6 Rydberg series attached to the ground hyperfine
states. The height of each stick is proportional to $1/\nu^3$ to indicate
the oscillator strength  available for each state. In calculating
these states, the $s$-Rydberg series have quantum defects near
4.32 plus 0.01 shifts depending on the channel and the $d$-Rydberg
series have quantum defects near 2.71 plus 0.01 shifts depending
on the channel. For these plots, there are two $d$-Rydberg series
attached to the $F_c=23/2$ and 21/2 threshold that are too close
together to distinguish in the plots.

Although there are more Rydberg series for $F_T=11$, these plots
already show quite complicated spectra. There are several places
where states attached to different threshold are nearly degenerate.
For example, near -971~GHz, two $d$-states attached to the $F_c=21/2$ threshold
are nearly degenerate with the $s$-series attached to the $F_c=23/2$
threshold.
As another example, there are 5 states in the 200~MHz range
between -404.47 and -404.27~GHz: 2 $d$-states attached to the
$F_c=23/2$ threshold, 2 $d$-states attached to the $F_c=21/2$
threshold and one $d$-state attached to the $F_c=19/2$ threshold.

\begin{figure}
\resizebox{80mm}{!}{\includegraphics{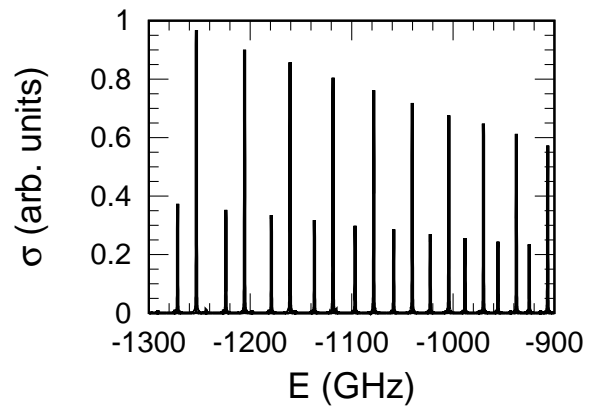}}
\caption{\label{FigCrs1}
A plot of the $d$-series photoabsorption cross section for $F_T=12$
using a linearly polarized photon with $M_{Ti}=1$. The taller series
of peaks is one of the $d$-series attached to the $F_c=23/2$ state
and the medium series is one of the series attached to the $F_c=21/2$ state.
The series attached to the $F_c=19/2$ threshold is $\sim 100$ times
smaller than the strong series.
}
\end{figure}

\begin{figure}
\resizebox{80mm}{!}{\includegraphics{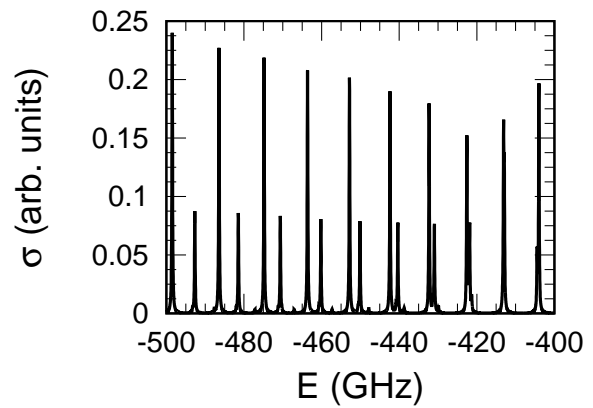}}
\caption{\label{FigCrs2}
Same as Fig.~\ref{FigCrs1}.
}
\end{figure}

The rest of the results we present include channel interactions caused by the hyperfine
splitting of the Ho$^+$.

Figures~\ref{FigCrs1} and \ref{FigCrs2} give an idea of what the
photoabsorption cross section would look like for the transition from the initial
state of Ref.~\cite{HPL} to the
$F_T=12$ series if only the $d$-series were present. For linearly polarized
light in the $z$-direction, the cross section is 0 if $M_T=0$ so we chose
$M_T=1$. We chose a linewidth of 50~MHz so the spectra would
not be a series of delta functions. We use the
frame transformation for the dipole matrix element as described in
Sec.~\ref{SecHoSig} and the K-matrix as described in Sec.~\ref{SecFT}
but with no corrections of the jj-coupled K-matrix.
We randomly assigned quantum defects to the LS-channels, the coupling
of Fig.~\ref{FigFT}b), in the
range seen by the experiment.\cite{HPL} The values we chose for
$\mu (SL)$ were: 2.85, 2.83, 2.81, 2.79 for $S=5/2$, $L=8-5$ and
2.77, 2.75, 2.73 for $S=3/2$, $L=8-6$.

One of the interesting features is that most of the states have little
oscillator strength. There are actually 5 Rydberg series with
$F_T=12$ and $\ell =2$ as discussed above, but only 2 of the series are
clearly visible in Fig.~\ref{FigCrs1}. The strongest series is one attached to the
$F_c=23/2$ threshold and the next strongest is one attached to
the $F_c=21/2$ threshold. None of the other 3 series has substantial
oscillator strength. The small oscillator strength leads to a much simpler
spectra compared to the actual energy levels. However, it must be
remembered that those states are still present and that external fields
or Rydberg-Rydberg interactions could cause strong mixing between these
nearly degenerate states.

\subsection{Zeeman Shift, Ho}

The effect of a weak magnetic field on the Rydberg series is deceptively
complicated due to the high density of states that can interact
through the magnetic field. For example, our simple calculation gives 15 $d$-Rydberg
levels in the 4~GHz region from $-489$ to $-485$~GHz with total
angular momentum $F_T\geq 10$. However, if the experiment can
restrict the states to high angular momentum, then the situation can be
favorable for isolating states. For example, if the Ho initial $F_{Ti}=12$
is in the $M_{Ti}=12$ state and then excited to the Rydberg states
with circularly polarized light then there are only two states with
$F_T=13$ close in energy: $-486.659$ and $-487.275$~GHz with
the first having more than 100 times the oscillator strength of the
second. These states can only mix with the $F_T=14,M_T=13$ state
at $-487.483$~GHz. Thus, if the $-486.659$~GHz state is excited, the
closest states are over 600~MHz away and will not strongly mix.
The closest $F_T=12$ states are $\simeq 210$~MHz away. Thus,
even an alignment mismatch that allows some $M_T=12$ character
will not have a large mixing with other states unless the Zeeman
shifts are above $\sim 20$~MHz in this example.

To understand how the energies shift with magnetic field,
we can treat the case where the magnetic field is in the
$z$-direction. In this case, the Zeeman shifts are given by
$\Delta E(M_T) = \mu B M_T$ where we can use the
Eqs.~(\ref{eqZeem1},\ref{eqZeem2}) to obtain an expression
for the $\mu$:
\begin{equation}
\mu= \frac{\langle \Phi_i||\mu_B(\ell^{(1)}+L_c^{(1)}+g_s[S_c^{(1)}+s^{(1)}])]-\mu_II^{(1)}||\Phi_i\rangle }
{\sqrt{(2F_T+1)(F_T+1)F_T}}
\end{equation}
where we have used an identity for the 3-j symbol when $F_T=F_T'$
and $q=0$.

For the two $F_T=13$ states, the $-486.659$~GHz state has
$\mu\simeq 0.85\mu_B$ and the $-487.275$~GHz has $0.92\mu_B$.
The coupling between them is $\sim 0.01\mu_B$. The $F_T=14$ state
at $-487.483$~GHz has $0.93\mu_B$. The $F_T=12$ states have
$1.02\mu_B$, $0.92\mu_B$, and $0.84\mu_B$ in order of increasing
energy. Thus, all of the states have a $\mu\sim\mu_B$ with a coupling
roughly two orders of magnitude smaller.

\subsection{Static Polarizability, Ho}\label{SecPol}

The polarizability determines the quadratic energy shift of a state
in an electric field, ${\cal E}$. The energy shift is $\Delta E=-(1/2)\alpha {\cal E}^2$.
Comparing to Eq.~(\ref{eqH2Ef}), the polarizability can be written
as $\alpha = -2H^{(2)}_{bb}|_{{\cal E}=1}$. Because the dipole
matrix elements scale like $\nu^2$ and the energy differences
scale like $1/\nu^3$, the polarizability scales as $\nu^7$.
For states with total angular momentum greater than 1/2, there is both
a scalar, $\alpha_0$, and tensor, $\alpha_2$, polarizability which captures
the $M$-dependence of the energy shift. If the electric field is in the
$z$-direction the change in energy is
\begin{equation}
\Delta E = -\frac{1}{2}\left(\alpha_0 + \frac{3 M^2 - F(F+1)}{F(2F-1)}\alpha_2\right){\cal E}^2
\end{equation}
where $F$ is the total angular momentum of the state and $M$ is
its projection on the $z$-direction.

The quantum defects for the $p$-Rydberg series are not known
very well but, from Refs.~\cite{STM,FTD}, they are expected to differ from
the quantum defects of the $d$-series by approximately 1. Thus,
the polarizability of the $d$-series can not be even qualitatively estimated
with current knowledge because slight changes in their quantum defects
could change the energy ordering of the states which would change
the sign of the polarizability. However, the magnitude of
the $d$-series polarizability should be relatively large due to the near
degeneracy.

As an example, we computed the static polarizability for the $s$-Rydberg
series with $F_T=12$. Reference~\cite{LWD} computed the frequency
dependent polarizability for the Ho ground level configuration.
From the discussion above, we expect that
this series does not have a rapidly varying quantum defect. In the
calculation, the quantum defect was fixed at $\mu_s=4.34$. Since
there is only one Rydberg series for this case, the energies are at
$-1/(2\nu_s^2)$ where $\nu_s = n-\mu_s$. For the
$p$-series, we chose the LS-coupled quantum defects, the coupling in
Fig.~\ref{FigFT}b), to be different
values between 3.73 and 3.83. For each Rydberg state, $n$, we used
all of the $p$-Rydberg states with $F_T=11,12,13$ that were between
energies $-1/[2 (\nu_s-6)^2]<E<-1/[2 (\nu_s+6)^2]$ and checked
that the results were converged by changing the energy range.

\begin{figure}
\resizebox{80mm}{!}{\includegraphics{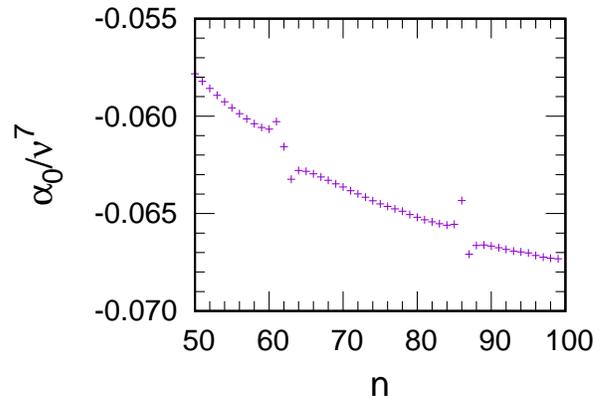}}
\caption{\label{FigPolvsn0}
The scaled static scalar polarizability, $\alpha_0 /\nu_s^7$, versus the principal quantum
number, $n$, for the $s$-Rydberg series with $F_T=12$.
}
\end{figure}

\begin{figure}
\resizebox{80mm}{!}{\includegraphics{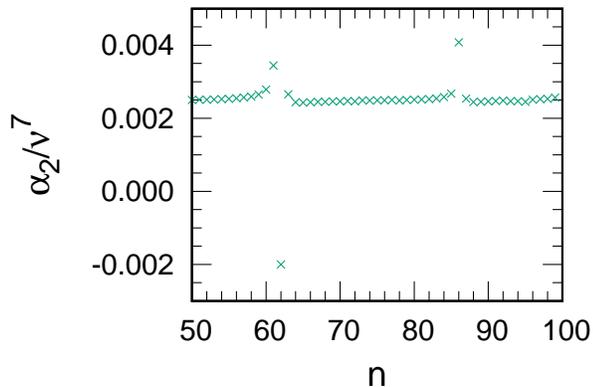}}
\caption{\label{FigPolvsn2}
The scaled static tensor polarizability, $\alpha_2 /\nu_s^7$, versus the principal quantum
number, $n$, for the $s$-Rydberg series with $F_T=12$.
}
\end{figure}

Figure~\ref{FigPolvsn0} shows the static scalar polarizability scaled by its main dependence
on $\nu$ and Fig.~\ref{FigPolvsn2} shows the static tensor polarizability also
scaled. The small magnitude of the tensor polarizability compared to the scalar
indicates that the variation of energy with $M_T$ is not large. The small relative
size of the tensor polarizability could be due to
the fact that most of the polarizability arises from an $s$-Rydberg electron
which would suppress the orientation dependence of the energy shift.
Over the range shown,
the scalar polarizability is negative which means the energy of a Rydberg state in
this series will shift up in energy with an increasing electric field. The size
of the scalar polarizability is relatively small because the $s$- and $p$-quantum
defects differ by approximately 0.5. This means the $p$-states nearly
evenly bracket each $s$-state which leads to the shift from each nearly
canceling each other.

Because there are $p$-series attached to the $F_c=21/2$ and 19/2
thresholds, there are cases where $p$-states are nearly degenerate
with an $s$-state. The effect of this can be seen near $n=62$ and 86.
Both the scalar and tensor polarizabilities have
a sharp variation near these cases. The variation
is not as large as might be expected because the near degeneracy means
the $p$-state wave function has a character that mostly consists of
the wrong core state. Thus, the dipole matrix element is smaller
than might be expected for the nearly degenerate
state. An interesting case is at $n=62$ where
the tensor polarizability changes sign. For most of the $n$, the energy
shift becomes smaller as $M_T$ increases, but the energy shift
increases with increasing $M_T$ at $n=62$.

For the real Ho atom, the $n$ where the resonance condition occurs
will probably be different from what was shown in this section. The
energy where the degeneracy occurs depends on the actual values of
the quantum defects. However, the number of regions where there
is a sharp variation should be $\simeq 2$ because that depends on
the threshold spacing which is well known.

\section{Two atom example results}\label{SecHoExTwo}

\subsection{$C_5$ coefficient, Ho}

We implemented the equations for the $C_5$ coefficients, Eq.~(\ref{eqC5}).
As an example, we calculated all of these coefficients for the $F_T=13$
state at $-486.659$~GHz discussed in the Zeeman shift section. The
states are labeled by the sum of the $z$-components of the total
angular momentum $M_T=M_{T1}+M_{T2}$. The number of states
at $M_T$ is ${\cal N}=2F_T-M_T+1$. The eigenstates are
even or odd with respect to interchange of the atoms. There is one more
even eigenstate than odd when ${\cal N}$ is odd, otherwise there are the
same number of even and odd states.

\begin{figure}
\resizebox{80mm}{!}{\includegraphics{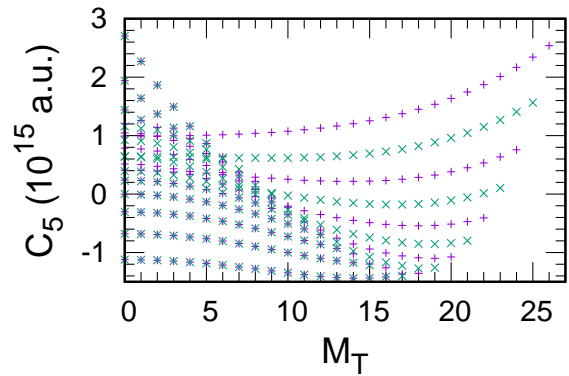}}
\caption{\label{FigC5}
The eigenvalues of the
$C_5$ coefficients as a function of  the total $z$-component of the
angular momenta, $M_T=M_{T1}+M_{T2}$. The even states (+)
and odd states (X) are plotted. The $C_5$ are in atomic units and scaled
by the factor $10^{15}$. To convert to GHz~$\mu$m$^5$, multiply
by $2.73\times 10^{-15}$.
}
\end{figure}

The results are plotted in Fig.~\ref{FigC5}. Because the $F_T=13$,
the maximum $z$-component of angular momentum is 26. The
states with negative $M_T$ are not plotted since the eigenvalues
do not depend on the sign of $M_T$. The overall size of the
$C_5$ should be $\sim \nu^8$ because there is a product of two
$r^2$ matrix elements, each of which scales like $\nu^4$. Because
this state is a mixture of Rydberg character with different thresholds,
there is a rough value $\nu\sim 82$ for this state
giving $C_5\sim 82^8\sim 2\times 10^{15}$.  The $C_5$ is roughly
this size. Converting to a frequency scale, the largest energy is $E\sim 7 GHz/R^5$
with $R$ in $\mu$m which suggests this interaction will not be important
in most applications.

There is an interesting pattern to
the eigenvalues. For large $M_T$, the even and odd eigenvalues
are quite distinct because the eigenvectors span all of the states
so that even and odd states have non-zero amplitude for states
with $M_{T1}\simeq M_{T2}$. As the $M_T$ becomes less than
$\sim 15$, an increasing number of states have nearly the same
eigenvalue for even and odd states. This is because these states
are mostly localized to large values of the difference in the
projections (i.e. large values of $|M_{T1}-M_{T2}|$). Since
these states have small amplitude for $M_{T1}\simeq M_{T2}$,
there is little amplitude to distinguish even from odd states.
These states are like a double well potential with a large barrier.

\subsection{$C_6$ coefficient, Ho}

As with the calculation of the polarizability, the $d$-Rydberg series
will be difficult to predict due to the near degeneracy from the $p$-Rydberg
series. So as with the polarizability, results are presented for the
$s$-Rydberg series with $F_T=12$ which can only mix with the
$p$-Rydberg series. As with the calculation of the
$C_5$ coefficient, there are even and odd eigenstates with respect to
interchange of atoms, with the number of states following the same
pattern as for the $C_5$.
The size of the $C_6$ coefficient is expected to scale with $\nu^{11}$
(four powers of dipole matrix element each scaling like $\nu^2$ divided
by an energy difference which scales like $\nu^{-3}$).

\begin{figure}
\resizebox{80mm}{!}{\includegraphics{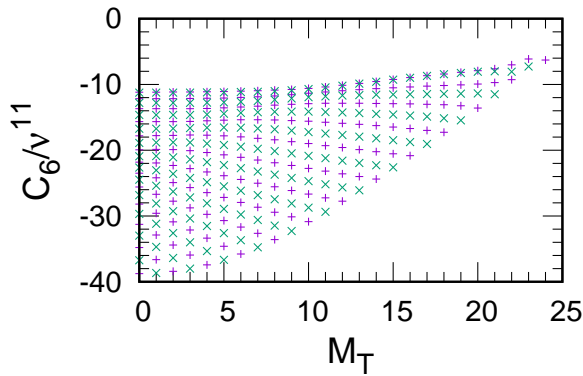}}
\caption{\label{FigC6vsMT}
The eigenvalues of the
scaled $C_6$ coefficients as a function of  the total $z$-component of the
angular momenta, $M_T=M_{T1}+M_{T2}$. The even states (+)
and odd states (X) are plotted. The case plotted is $n=50$. To convert
to GHz~$\mu$m$^6$, multiply
by $1.44\times 10^{-19}$.
}
\end{figure}

Figure~\ref{FigC6vsMT} shows the $M_T$ dependence of the eigenvalues
of the $C_6$ matrix scaled by the expected $\nu$-dependence. The state
plotted is $n =50$ which is far from the cases that are sensitive to $n$.
As with the $C_5$ eigenvalues, there is an interesting pattern to the
even and odd eigenvalues. Unlike the $C_5$ case, the even and odd
values are distinct except for the eigenvalue with smallest magnitude.
All of the eigenvalues are negative which leads to an attractive potential
between the atoms independent of the $M_T$. The size of the $C_6$
coefficients spans a wide range of values: over a factor of 6 from the smallest
to largest in magnitude. The overall size of the van der Waals interaction for this series is not
especially strong. At $n=50$, the smallest magnitude $C_6$ is approximately
a factor of $\sim 20$ smaller than that for Rb $50d_{5/2}50d_{5/2}$ with $M_T=5$
while the largest magnitude Ho $C_6$ is a factor of $\sim 3$ times smaller.

\begin{figure}
\resizebox{80mm}{!}{\includegraphics{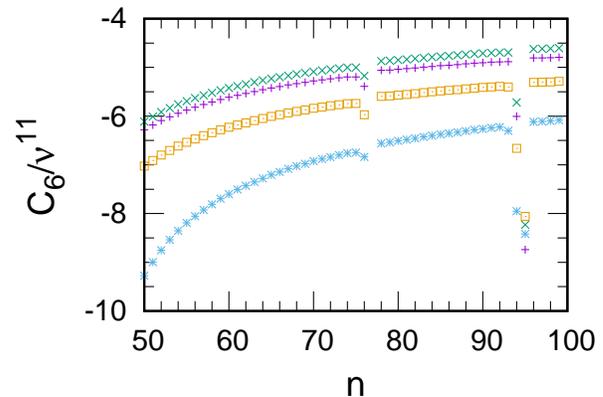}}
\caption{\label{FigC6vsn}
The eigenvalues of the
scaled $C_6$ coefficients as a function of the principal
quantum number $n$. The $M_T=24$ even states (+),
$M_T=23$ even states (X), and the two $M_T=22$ even
states (square and asterisk) are plotted.  To convert
to GHz~$\mu$m$^6$, multiply
by $1.44\times 10^{-19}$.
}
\end{figure}

It is difficult to show the dependence of the $C_6$ on $n$ for all of
the possible $M_T$. In Fig.~\ref{FigC6vsn}, the scaled $C_6$ for the four even states
with largest $M_T$ are shown. As with the polarizability, there are two
places ($n\sim77$ and 95)
where the $C_6$ varies rapidly with $n$. The values for $n=77$ are
not shown because they are a factor of $\sim -20$ of the average
value. For these states, only the $n=77$ leads to a positive $C_6$
which gives a repelling potential between the two atoms.
The actual $C_6$ near these sensitive $n$ can not be predicted with
the current state of knowledge of Ho. However, it is likely that there
will be cases of strong $C_6$.

\section{Summary}

We have derived the equations that can be used to treat the Rydberg
states of atoms where the core state has sizable hyperfine splitting.
This could be interesting for rare earth atoms where the ground state
of the ion can have very large angular momentum as well as many
hyperfine levels.

The theory is developed using the tools of multichannel quantum defect
theory (MQDT). We have derived the equations for both single atom
properties and two atom properties. For a single atom, we have given
expressions for finding the energy levels, oscillator strengths,
Zeeman mixing and shifts, and Stark mixing and shifts. For an atom
pair, we have shown how to calculate the Rydberg-Rydberg interactions
in general and have derived the specific cases for the $C_5$ and
$C_6$ coefficients.

Although the treatment above should be accurate enough for many
applications, the theory needs substantial input from measurements
of the single atom properties. This might be a challenge for many
atoms. Although the Rydberg states are not known well enough
for any of the rare earths, we made estimates of parameters for Ho
and used the estimates to demonstrate how to implement the
equations for both single atom and two atom parameters. These
results give a cartoon picture how the parameters might behave
in a real atom.

\section{acknowledgment}
This work was supported by the National Science Foundation
under award No.1404419-PHY (FR) and award No.1707854-PHY (DWB and MS).

\bibliography{holmium_theory}

\end{document}